\documentclass{jaa}
\usepackage{newtxtext,newtxmath}
\usepackage[T1]{fontenc}
\usepackage{ae,aecompl}
\usepackage{graphicx}
\usepackage{caption}
\usepackage{subcaption}
\usepackage[utf8x]{inputenc}
\usepackage{gensymb}
\usepackage{float}
\usepackage{hyperref}

\begin{document}\sloppy

\title{UOCS. V. UV study of the old open cluster NGC 188 using AstroSat}


\author{Sharmila Rani\textsuperscript{1,2,*}, Annapurni Subramaniam\textsuperscript{1}, Sindhu Pandey\textsuperscript{3}, Snehalata Sahu\textsuperscript{1}, Chayan Mondal\textsuperscript{1}, Gajendra Pandey\textsuperscript{1}}
\affilOne{\textsuperscript{1}Indian Institute of Astrophysics, Koramangala II Block, Bengaluru, 560034, India\\}
\affilTwo{\textsuperscript{2}Pondicherry University, R.V. Nagar, Kalapet, 605014, Puducherry, India\\}
\affilThree{\textsuperscript{3} Aryabhatta Research Institute of Observational Sciences (ARIES), Manora Peak, Nainital, 263001, India}

\twocolumn[{

\maketitle

\corres{sharmila.rani@iiap.res.in}

\msinfo{XXX}{YYY}

\begin{abstract}
We present the UV photometry of the old open cluster NGC188 obtained using images acquired with Ultraviolet Imaging telescope (UVIT) on board the \textit{ASTROSAT} satellite, in two far-UV (FUV) and one near-UV (NUV) filters. 
UVIT data is utilised in combination with optical photometric data to construct the optical and UV colour-magnitude diagrams (CMDs). In the FUV images, we detect only hot and bright blue straggler stars (BSSs), one hot subdwarf, and one white dwarf (WD) candidate. In the NUV images, we detect members up to a faintness limit of $\sim22$ mag including 21 BSSs, 2 yellow straggler stars (YSSs), and one WD candidate. This study presents the first NUV-optical CMDs, and are overlaid with updated BaSTI-IAC isochrones and WD cooling sequence, which are found to fit well to the observed CMDs.  We use spectral energy distribution (SED) fitting to estimate the effective temperatures, radii, and luminosities of the UV-bright stars. We find the cluster to have an HB population with three stars (T$_{eff}$ = 4750k - 21000K). We also detect two yellow straggler stars, with one of them with UV excess connected to its binarity and X-ray emission. 
\end{abstract}


}]


\doinum{12.3456/s78910-011-012-3}
\artcitid{\#\#\#\#}
\volnum{000}
\year{0000}
\pgrange{1--}
\setcounter{page}{1}
\lp{1}

\section{Introduction}
Old open clusters (OC) provide the ideal environments for studying the formation and evolution of single and binary stellar populations. OCs are extensively used to understand the evolution of the Galactic disk, and chemical and dynamical evolution of the Galaxy. Most of the OCs are studied in the optical band, but only a handful of them are studied in the Ultraviolet (UV) band; this is due to the lack of space-based UV telescopes, which has improved during recent times. Looking into the properties of OCs in the UV region is crucial in understanding the cluster's hotter stellar populations, which emit a significant fraction of flux in the UV domain; hence, an essential tool in discovering hot stars. The UV light of stellar populations is contributed by exotic populations such as Blue Straggler stars (BSSs), Horizontal branch (HB) stars, hot White Dwarfs (WDs), Subdwarfs, cataclysmic variables, etc. All the stars mentioned above are detectable in optical and infrared bands, but as these stars are hot, the bolometric corrections make them optically faint and indistinguishable from cooler stars. In the old OCs M67, NGC 188, and NGC 6791, hot stars have been detected and studied by Landsman {\em et al.} (1998) using \textit{Ultraviolet Imaging Telescope (UIT)} data. De Martino {\em et al.} (2008) have obtained FUV and NUV images of the old OC NGC 2420 from \textit{Galaxy Evolution Explorer (GALEX)} and cross-matched with Sloan Digital Sky Survey (SDSS) u,g,r,i,z photometric data in search of WDs in the cluster. The wide field photometry for nearby clusters M67, NGC 188, NGC 2539, and M79 using \textit{Swift Ultraviolet Optical Telescope (UVOT)} by Siegel {\em et al.} (2014) shows that the UV colour-magnitude diagrams (CMDs) can easily identify the unusual UV bright stars. Browne {\em et al.} (2009) used \textit{GALEX} observations to check UV variability in OCs Hyades and the Pleiades and detected 16 UV variable sources. Gosnell {\em et al.} (2015) utilised \textit{Hubble Space Telescope (HST)} data in far-ultraviolet (FUV) passbands to detect WD companions to BSSs in old OC NGC 188. Thus, UV imaging of OCs provides a unique window for identifying and characterising hot stellar populations to further shed light on their formation and evolution.\\

The importance of studying OCs in the UV, as outlined above, has lead to the formation of the Ultra Violet Imaging Telescope (UVIT) OC study (UOCS), as described in Jadhav {\em et al.}(2019). The program was initiated to understand the properties of single and binary stars in OCs.\\ 

BSSs are defined as the stars that are observed to be brighter and bluer than the corresponding main-sequence (MS) turnoff in an optical CMD in star clusters (Sandage 1953). The origin of these stars is unexplained by the standard theory of single-star evolution, but they are found to be confirmed members of the clusters. The primary scenarios proposed to explain the formation of BSSs in clusters are mass transfer in close binary systems ( McCrea {\em et al.} 1964) and stellar mergers resulting from direct stellar collisions ( Hills \& Day {\em et al.} 1976). Yellow Straggler stars (YSSs) are brighter than stars in the sub-giant branch (SGB), and bluer than the red giant branch (RGB). Observationally, they are found to lie between blue straggler and RGB regions on the CMD ( See Sindhu {\em et al.} 2018 and references therein). These stars are also described as evolved BSSs.  The hot subdwarfs of B and O-type (sdB, sdO) represent the late stages of the evolution of low mass stars. sdB stars are considered as the metal-rich counterparts of the extreme HB (EHB) stars in globular clusters (GCs). Till now, a few hot subdwarfs and HB stars are detected in two metal-rich and old OCs, i.e., NGC 6791 and NGC 188 (Kaluzny \& Udalski 1992; Liebert {\em et al.} 1994; Green {\em et al.} 1997). Only one sdB star is found in NGC 188, whereas NGC 6791 hosts five sdB stars. Schindler {\em et al. }(2015) investigated 15 OCs to understand formation of EHB stars by taking NGC 188 and NGC 6791 as template clusters. They identified only red giant clump stars but no EHB stars were detected.\\

In this paper, we study one of the oldest  and well studied OC, NGC 188, known in the Galaxy. The cluster is studied widely due to its richness, metallicity, age and its location in the galactic plane ($\alpha_{2000} = 0^{h}47^{m}12^{s}.5$, $\delta_{2000} = +85\degree14{'}49{''}$, \textit{l} = $122\degree.85$, \textit{b} = $+22\degree.38$). The age of this cluster is determined to be 7 Gyr (Sarajedini {\em et al.} 1999), and the reddening of the cluster is 0.036 $\pm$ 0.01 mag (Jiaxin {\em et al.} 2015). This cluster is located at a distance of 1866 pc (Gao 2018) and the metallicity is found to be solar (Sarajedini {\em et al.} 1999). In this study, we present the results of the UV imaging of the NGC 188 in two FUV and one NUV filters using an ultraviolet imaging telescope (UVIT) on {\it AstroSat}. We characterise the UV bright stars identified in this cluster by analysing SEDs to throw light on their formation and evolution.\\

The paper is organised as follows: Section~\ref{sec:2} describes the observations and data analysis. In Section~\ref{sec:3}, Optical, and UV CMDs, along with the results obtained from SED analysis, are described. All the results are discussed in Section~\ref{sec:4} The results are summarised in Section~\ref{sec:5}

\begin{figure}
    \centering
	\includegraphics[scale=0.46]{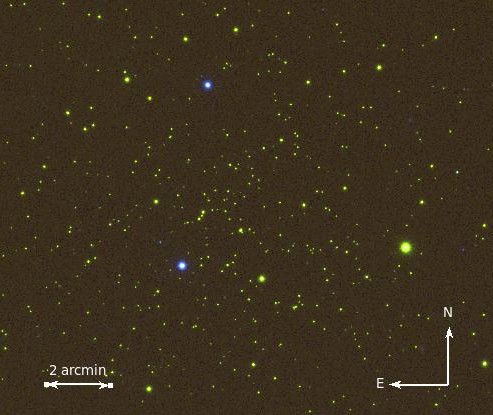}
    \caption{UVIT image of NGC 188 obtained by combining images in NUV (N279N) and FUV (F148W) channels. Yellow and blue colour corresponds to NUV and FUV detections, respectively.}
    \label{combimage}
\end{figure}

\section{Observations and Data Analysis}
\label{sec:2}
The data used in this study were acquired with the UVIT instrument on-board the \textit{AstroSat} satellite. UVIT is one of the payloads in {\it AstroSat}, the first Indian multi-wavelength space observatory. The UVIT instrument has two 38 cm telescopes: one observes only in FUV region ($\lambda = 130-180$ nm), and the other in NUV ($\lambda = 200-300$ nm) and VIS ($\lambda = 320-550$ nm) region. UVIT can simultaneously generate images in the FUV, NUV and VIS channels in a circular field of view of $\sim28'$ diameter. Every channel consists of multiple filters covering a different wavelength range. The data obtained in VIS passband is used for drift correcting the images. The detectors used for the UV and VIS passbands operate in photon counting mode and integration mode, respectively. Further details about UVIT instrument can be found in Tandon {\em et al.} (2017a) and Tandon {\em et al.} (2020), along with calibration results. The magnitude system adopted for UVIT filters is similar to the used for GALEX filters, and hence the estimated magnitudes will be in the AB magnitude system.\\
\begin{figure}
   \hspace*{-0.7cm} 
	\includegraphics[height=9.5cm, width=10cm]{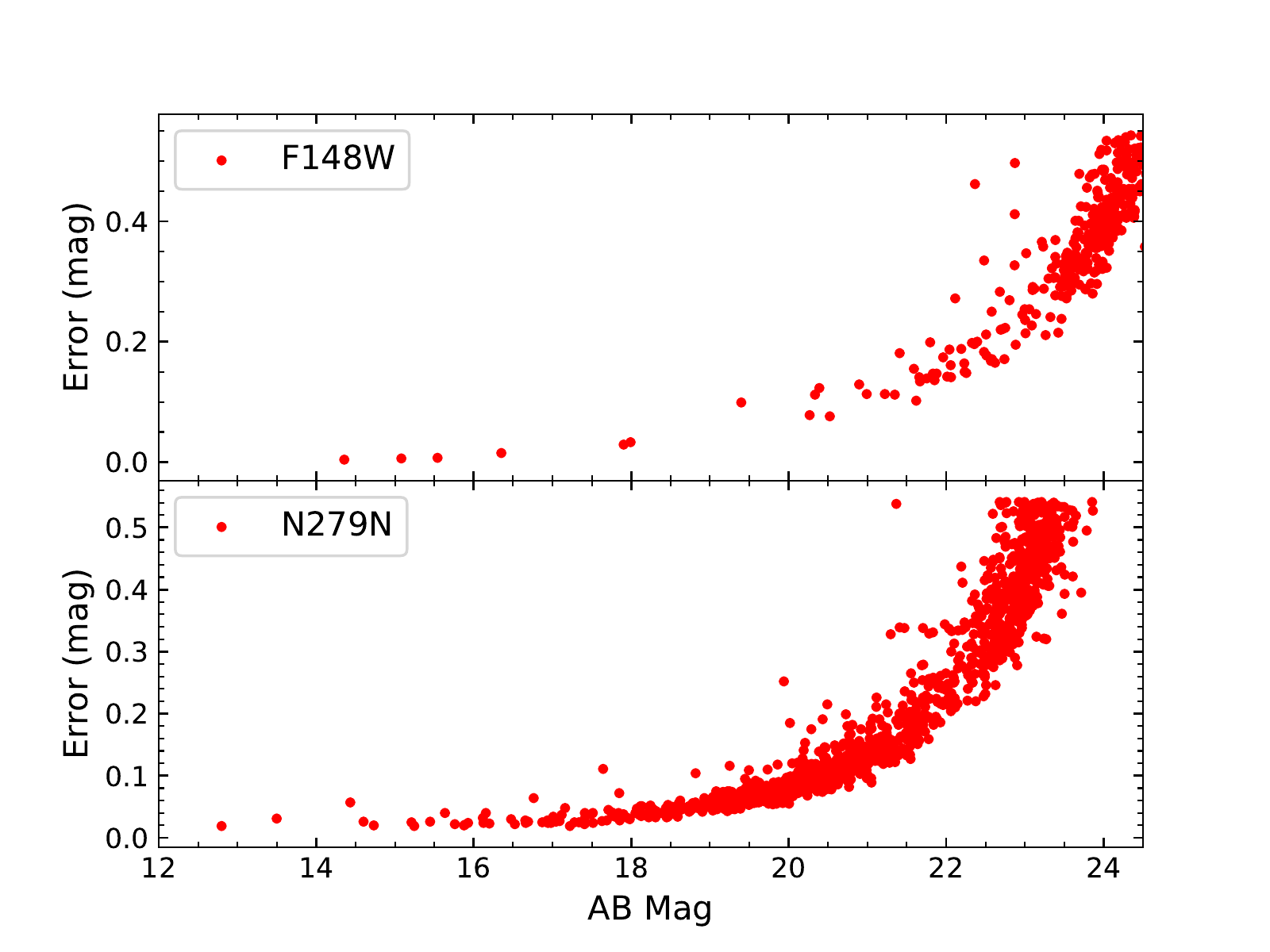}
     \caption{PSF fit errors as a function of magnitude for the UVIT observations of NGC 188. The top panel shows the errors for FUV F148W filter, whereas the bottom panel shows for the NUV N279N filter.}
    \label{photoerr}
\end{figure}

NGC 188 was observed as UVIT’s ”first light” object on 30 November 2015. The cluster was observed every month to track the variation in UVIT senstivity over the first six months of its operation. The observations of NGC 188 were done from December 2015 to March 2016. All data used in the analysis were taken in three filters: one NUV (N279N) and two FUV (F148W and F172M). Images were corrected for distortion, flat field illumination, and spacecraft drift utilising the customised software package CCDLAB (Postma \& Leahy 2017). The images created for each orbit were then aligned and combined to produce a science ready image. The UVIT observational details of the NGC 188 UV images are tabulated in Table~\ref{tab:1}. The UVIT image of the cluster created using one FUV F148W and one NUV N279N filter is shown in Figure~\ref{combimage} with blue and yellow corresponding to FUV and NUV detections, respectively.

\subsection{Photometry}
\label{sec:2.1}
Aperture photometry was carried out on the FUV F172M image to get the estimate of counts. 
We performed psf photometry on one FUV F148W and NUV N279N images using the IRAF/NOAO package DAOPHOT (Stetson 1987). A curve of growth technique was used to determine the aperture correction value and then applied to all magnitudes estimated using aperture and psf photometry. To obtain final magnitudes in each filter, saturation correction was applied according to the method described in Tandon {\em et al.} (2017b). The instrumental magnitudes are calibrated to the AB magnitude system using zero-point (ZP) magnitudes reported in  Tandon {\em et al.} (2017b). The PSF fit errors in estimated magnitudes are shown in Figure~\ref{photoerr} as a function of magnitude for FUV and NUV filters. The stars as faint as 22 mag are detected both in FUV and NUV images with typical error 0.2 mag and 0.3 mag, respectively. The observed UVIT magnitudes are corrected for extinction. In order to compute the extinction value in visible band ($A_V$), We have adopted reddening E(B$-$V) = 0.036 mag (Jiaxin {\em et al.} 2015) and the ratio of total-to-selective extinction as $R_v$ = 3.1 from Whitford (1958) for the Milky Way. The Fitzpatrick reddening relation (Fitzpatrick 1999) is used to determine extinction co-efficients $A_\lambda$ for all bandpasses.
\begin{table}
\centering
	\caption{Details of observations of NGC188}
	\label{tab:1}
	\makebox[0.7\linewidth]
	{
	\begin{tabular}{cccccc} 
		\hline
		\hline
		 Filter & $\lambda_{mean}$ & $\Delta\lambda$  & ZP  & Exposure \\
		  & ({\AA}) & ({\AA}) & (AB mag) & Time (sec) \\
		\hline
		 F148W & 1481 & 500 & 18.016 & 3216 \\
		 F172M & 1717 & 125 & 16.342 & 4708 \\
		 N279N & 2792 & 90  & 16.50 & 9365 \\
		\hline
	\end{tabular}
	}
\end{table}
\begin{table*}[t]
\centering
\caption{Parameters of NGC 188 used in this study}
\label{ngc188params}
\makebox[0.9\linewidth]
{
\begin{tabular}{ccc}
\hline
\hline
Parameters & Value & Reference \\
\hline
Metallicity (Z) & 0.02 & Jiaxin {\em et al.} (2015)\\
Age    & 7 $\pm$ 0.5 Gyr & Sarajedini {\em et al.} (1999) \\
Distance modulus, ${\it (m-M)_{V}}$ & 11.44 $\pm$ 0.08 mag & Sarajedini {\em et al.} (1999) \\
Reddening, {\it E(B$-$V)} & 0.036 $\pm$ 0.01 mag & Jiaxin {\em et al.} (2015) \\
\hline
\end{tabular}
}
\end{table*}

\begin{figure*}[htb]
\centering
\begin{subfigure}{0.5\textwidth}
  \centering
  \includegraphics[height=8.0cm,width=9.0cm]{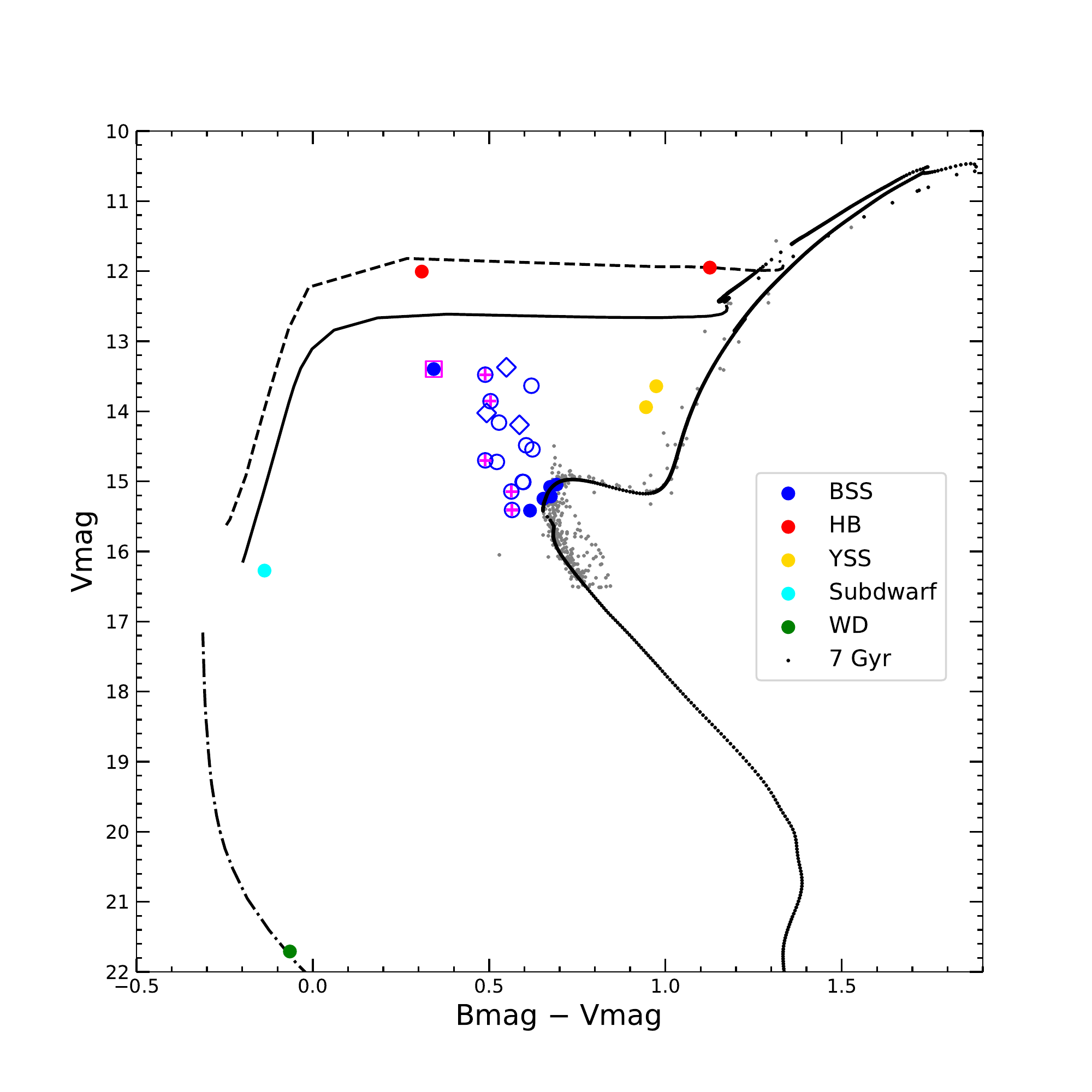}
  \end{subfigure}%
\begin{subfigure}{0.5\textwidth}
  \centering
  \includegraphics[height=8.0cm,width=9.0cm]{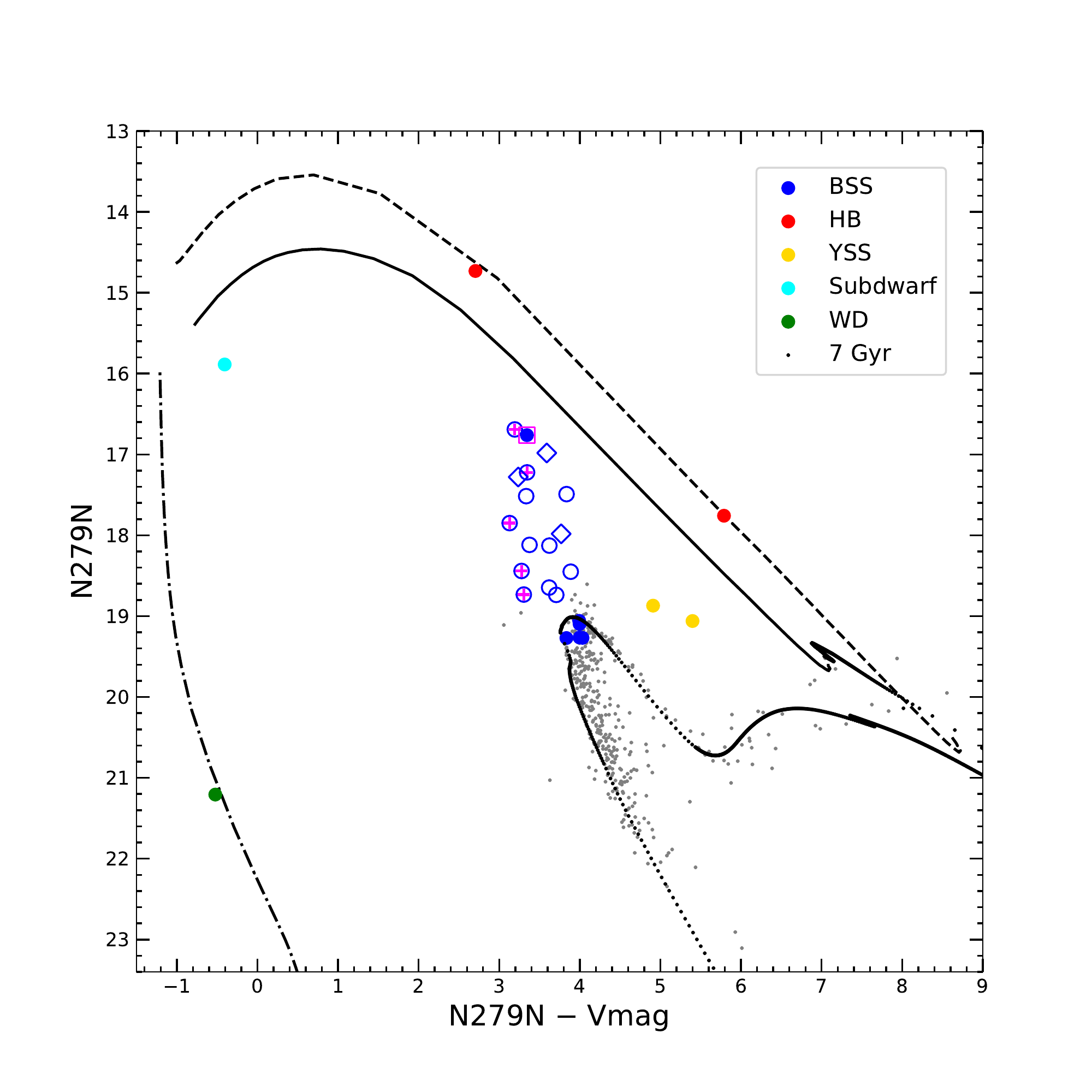}
\end{subfigure}
\caption{(Left panel) Optical CMD of NGC 188 of all member stars co-detected with UVIT N279N filter and optical photometric data. (Right panel) NUV-optical CMD of NGC 188 of member stars cross-identified using UVIT N279N data with optical photometric data. The meaning of all the symbols are marked in the figures. Previously known binary and single BSSs are shown as open blue circles and open blue diamonds, respectively. The BSSs with WD detections are shown with a magenta plus symbol. The BSS outlined with magenta square symbol is bright in both NUV and FUV CMDs. The over-plotted black colour dots represent updated BaSTI-IAC model isochrones generated for an age 7 Gyr and a solar metallicity. The solid and dashed lines shown along the HB track correspond to zero-age HB (ZAHB) and terminal-age HB (TAHB), respectively. The dashed-dotted black line indicates the WD cooling sequence for a WD with mass 0.5M$_{\odot}$.}
\label{nuvcmds}
\end{figure*}
\section{Results}
\label{sec:3}

\subsection{UV and Optical Colour Magnitude Diagrams}
\label{sec:3.1}
\begin{figure*}[t]
\centering
\begin{subfigure}{0.5\textwidth}
  \centering
  \includegraphics[height=8.0cm,width=9.0cm]{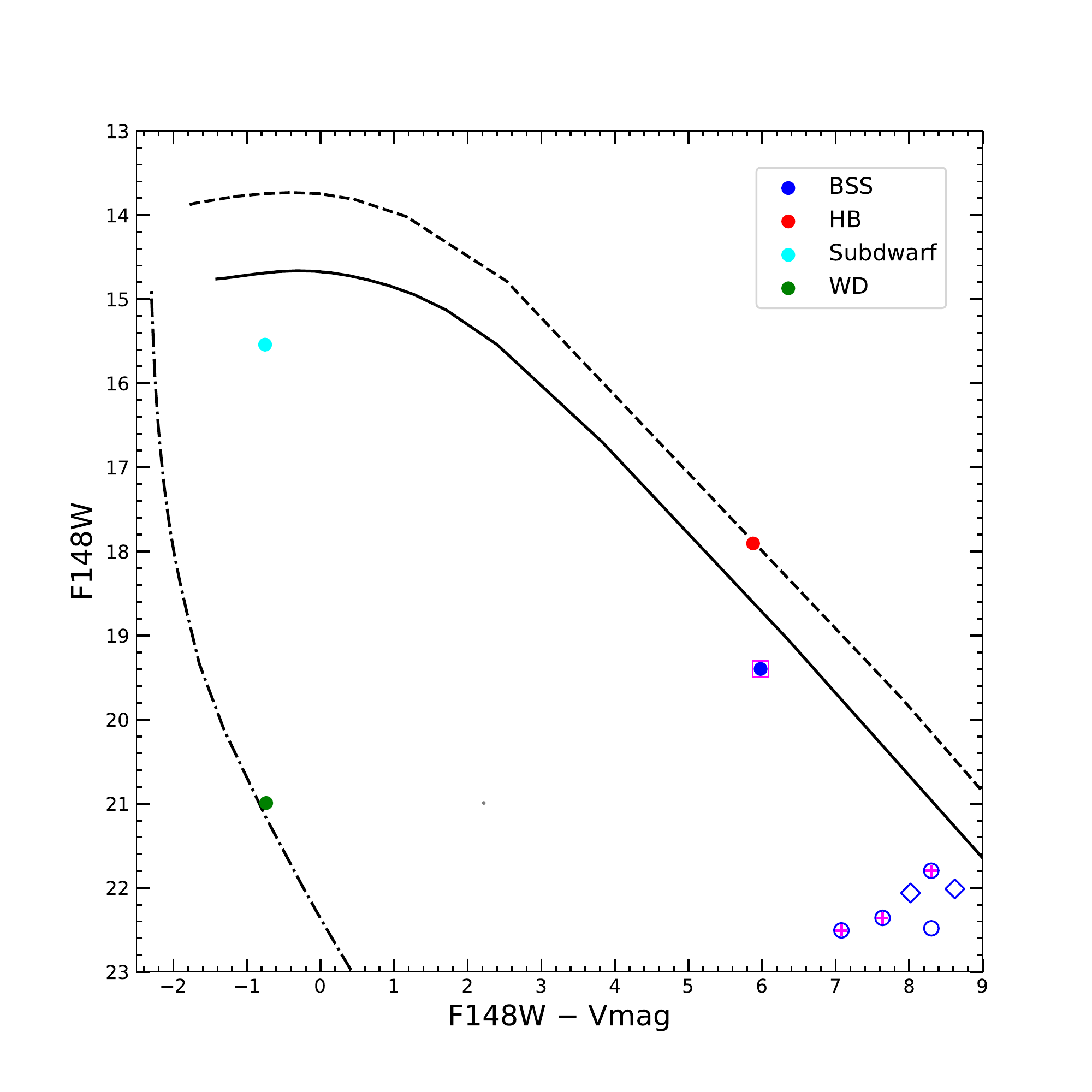}
  \end{subfigure}%
\begin{subfigure}{0.5\textwidth}
  \centering
  \includegraphics[height=8.0cm,width=9.0cm]{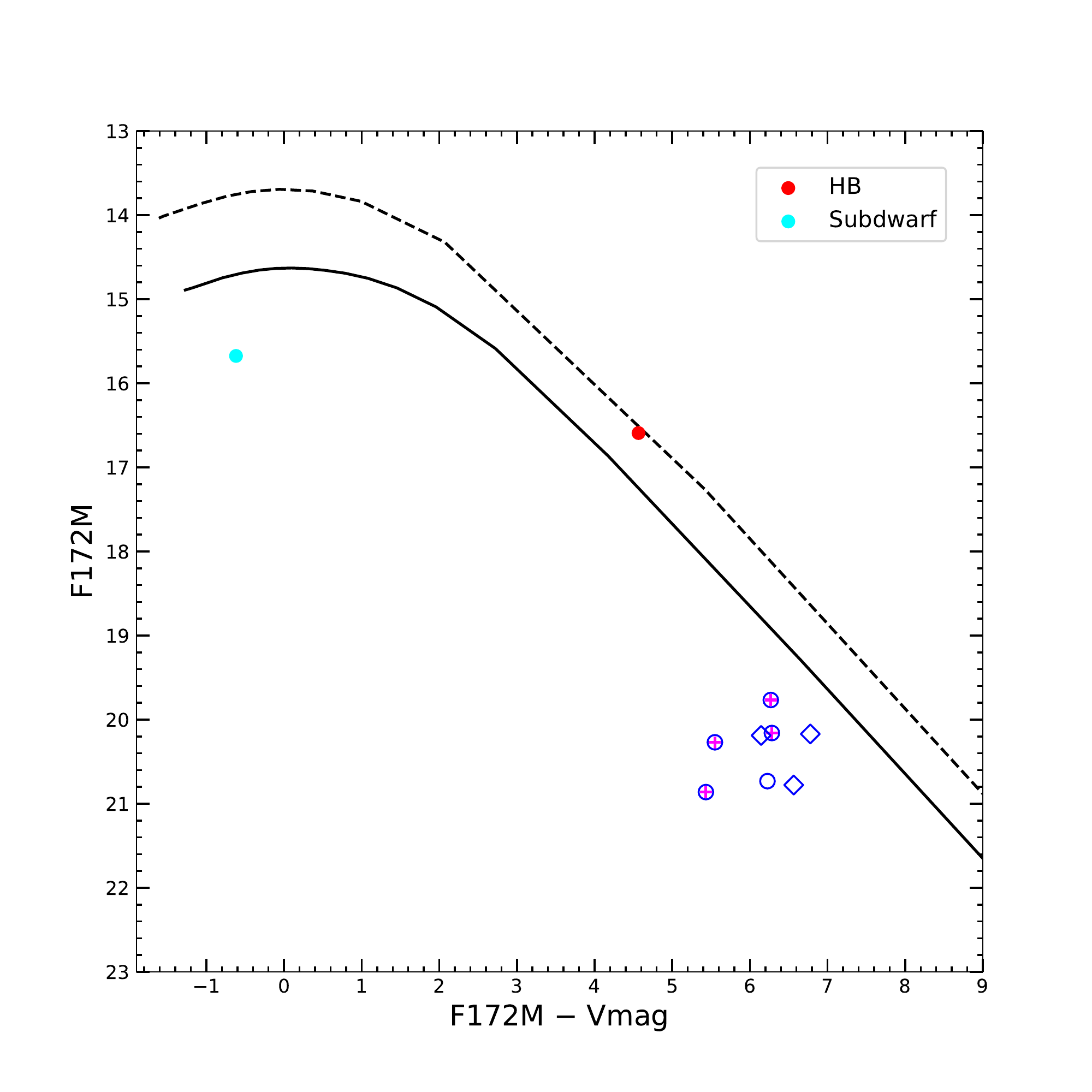}
\end{subfigure}
\caption{FUV-optical CMDs of NGC 188 of member stars cross-identified using UVIT FUV and ground-based optical photometric data. Other details are same as in Figure~\ref{nuvcmds}}
\label{fuvcmds}
\end{figure*}
To identify and classify the stars detected with UVIT into various evolutionary phases, We have used ground-based optical photometric data (Sarajedini {\em et al.} 1999) of NGC 188 to cross-identify with the UVIT detected stellar sources. First, we have selected stars with proper motion (PM) membership probability more than 50$\%$ as most likely members of the cluster from the catalog given by  Platais {\em et al.} (2003), which gives us 562 member stars. Then we have cross-matched optical photometric data taken from Sarajedini {\em et al.} (1999) with PM catalog. Geller {\em et al.} (2008) presented the results of radial-velocity (RV) survey of NGC 188 using WIYN data. They measured the radial velocities for 1046 stars in the direction of NGC 188, and further calculated the RV membership probability for all the stars. Out of 1046 stars, 473 stars are found to be likely cluster members. Further, we cross-matched optical data with RV catalog to check for the RV membership probability. After choosing all the PM and RV members in the cluster, we further cross-matched with UVIT detected stars in all the three filters. We have also included the stars for which PM membership probability is given, but RV membership is not known. We have cross-identified 356 stars of NGC 188, as a cluster members, in NUV passband and 10 cluster members in both FUV passbands. The accuracy of this cross-match is within $1.5''$. 24 BSS candidates have been cataloged by Ahumada \& Lapasset (2007). Geller {\em et al.} (2008), identified 20 BSSs to be confirmed members based on the radial velocity membership.\\ 

As there are many stars detected in the NUV, the detected members are segregated based on their detection in the NUV.  In the NUV filter, we detect the MS, turn-off, SGB, RGB, BS and YS stars, whereas in FUV images, only hot and bright stars are detected. We have detected 21 BSSs previously known in literature, out of which sixteen are RV members, and five are PM members. One previously known hot subdwarf is also detected in both FUV and NUV images, but RV membership for this star is not known as it is fainter than 16 mag in optical CMD. Two YSSs are selected on the basis of their position in the optical CMD. We have also detected one very bright and hot star in NUV and one FUV image, probably a WD candidate. Both PM and RV memberships are unknown for this star. For only cross-matched member stars, we have created the optical and UV-optical CMDs as shown in Figures~\ref{nuvcmds} and \ref{fuvcmds}. The parameters adopted in this study are listed in Table~\ref{ngc188params}. The optical and UV-optical CMDs are over-plotted with updated BaSTI-IAC isochrones (Hidalgo {\em et al.} 2018). The updated BaSTI-IAC\footnote{\url{http://basti-iac.oa-abruzzo.inaf.it/}} isochrones are generated for an age 7 Gyr, a distance modulus of 11.44 mag (Sarajedini {\em et al.} 1999) and, a solar metallicity with helium abundance Y = 0.247, \big[$\alpha$/H\big] = 0, including overshooting, diffusion, and mass loss efficiency parameter $\eta$ = 0.3. The BaSTI-IAC model also provides HB model, which includes zero age HB (ZAHB), post-ZAHB tracks and end of the He phase known as terminal age HB (TAHB) with or without diffusion for a selective mass range. We generated the ZAHB and TAHB tracks for a solar metallicity including diffusion.\\ 

The overlaid BaSTI-IAC isochrones are fitting well to the observations in both optical and NUV-optical CMDs as shown in Figure~\ref{nuvcmds}. This is probably the first UV CMD for this well studied cluster in the NUV, the overlaid isochrones match the observed sequence more or less satisfactorily. The CMDs shown in figure~\ref{nuvcmds}, suggests that this cluster probably has a HB population, with stars in the HB. Two stars marked with red colour are lying along HB tracks, implying that these stars may belong to the HB evolutionary phase. In fact, they are lying close to TAHB track, indicating that they are about to evolve off the HB phase. The previously known subdwarf is found close to the blue end of the ZAHB track in the optical CMD, and fainter than the blue end in UV CMDs. 
We have also shown a WD cooling curve for a $0.5M_{\odot}$ WD with hydrogen envelope with dashed-dotted black line in Figures~\ref{nuvcmds} and \ref{fuvcmds} provided by Tremblay {\em et al.} (2011) to define the location of WDs in all CMDs (P. Bergeron, private communication). Note that star marked with green colour in all CMDs is lying along the WD cooling track, indicating that the star is a possible WD candidate, although its membership is uncertain. We have also detected two stars which are bright in NUV CMDs with respect to other giants shown with yellow colour in the optical as well as in NUV-optical CMD (See Figure~\ref{nuvcmds}). These two stars are bluer than the RGB and brighter than SGB track in the optical CMD, but they are bright in the NUV CMD. These two stars are classified as YSSs.
\begin{figure*}[t]
\centering
\begin{subfigure}{0.5\textwidth}
  \centering
  \includegraphics[height=8.0cm,width=9.0cm]{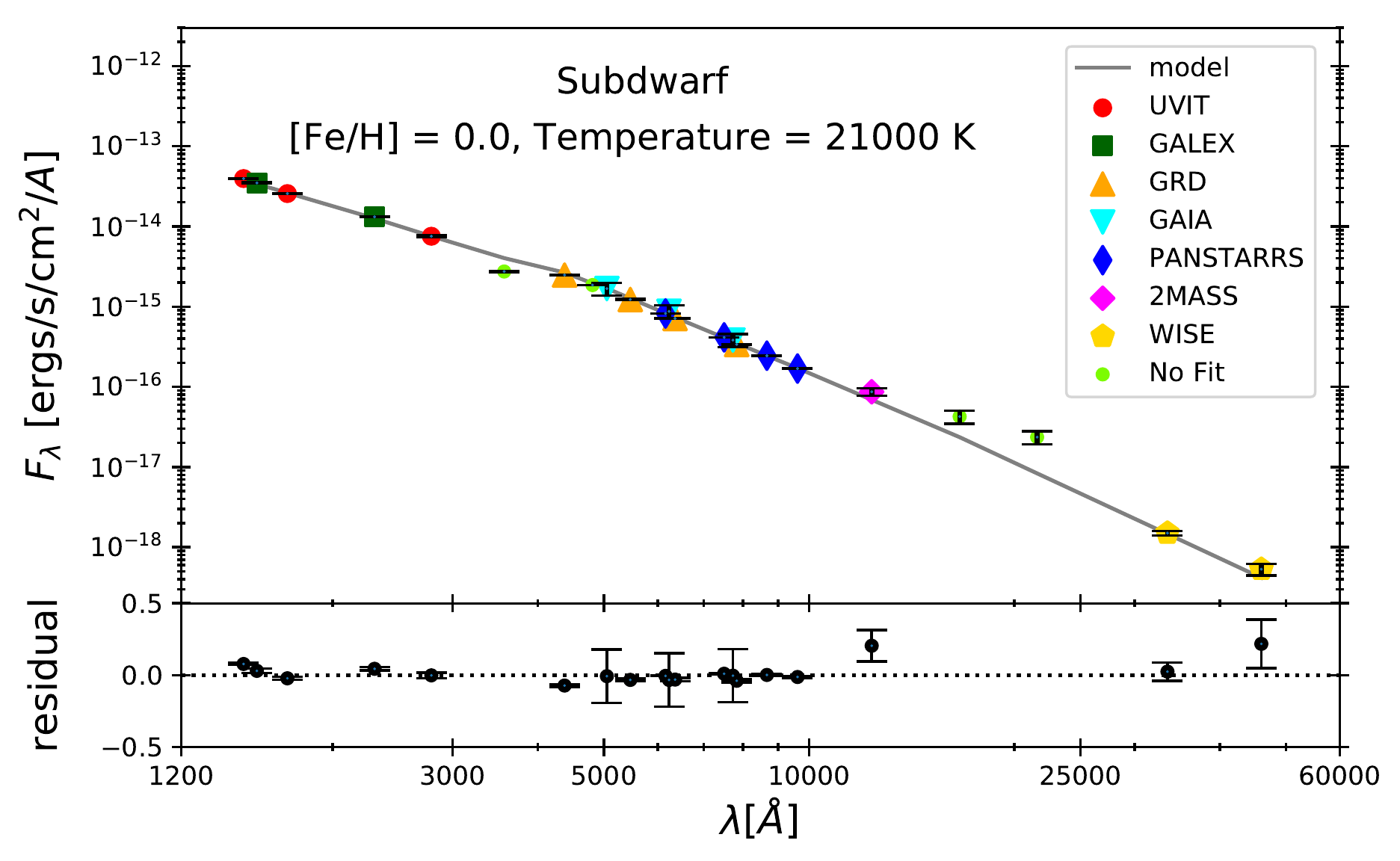}
  \end{subfigure}%
\begin{subfigure}{0.5\textwidth}
  \centering
  \includegraphics[height=8.0cm,width=9.0cm]{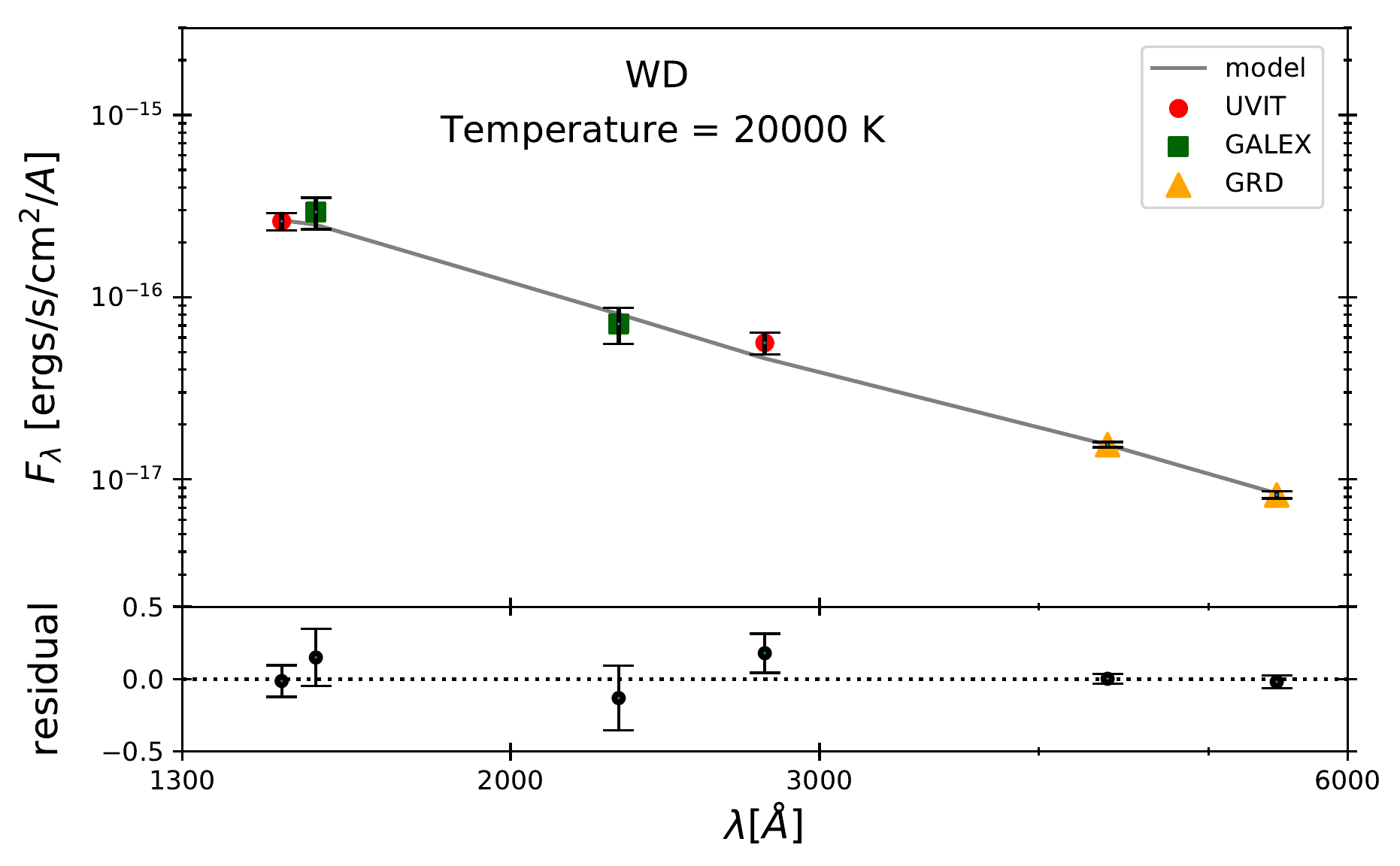}
\end{subfigure}
\caption{Spectral energy distribution (SED) of a hot subdwarf and WD candidate detected with UVIT after correcting for extinction. The best fit parameters are shown in the figure.}
\label{seds}
\end{figure*}

\subsection{Spectral Energy Distributions of UV bright stars}
\label{sec:3.2}
In this section, our main goal is to check the evolutionary status of stars, which appear bright in NUV and FUV CMDs, 
by determining their stellar parameters using spectral energy distribution (SED) fit technique. We constructed the SEDs to estimate the stellar parameters such as effective temperature ($T_{eff}$), luminosity (L), and radius (R) of the stars which appear bright and hot in the UV CMDs. To create the SEDs, We have used observed photometric data points spanning a wavelength range from FUV-to-Infrared and then fitted with selected theoretical models. The virtual observatory tool, VOSA (VO Sed Analyser, Bayo {\em et al.} 2008) is used for SED analysis. The filter transmission curves are employed to calculate the synthetic photometry of the selected theoretical models. VOSA utilises the fixed distance to the cluster to scale the synthetic fluxes with the observed fluxes. After constructing the synthetic SED, a $\chi^2$ minimisation test is performed to compare the observed with the synthetic photometry to find the best fit parameters of the SED. The reduced $\chi^2_{red}$ value for the best fit is evaluated using the expression given below:

\begin{equation*}
\hspace*{1.0cm}
   \chi^2_{red} = \frac{1}{N-N_{f}} \displaystyle \sum_{i=1}^{N} {\frac{(F_{o,i} - M_{d}F_{m,i})^2}{\sigma_{o,i}^2}}
\end{equation*}

where N is the number of photometric data points, $N_{f}$ is the number of free parameters in the model, $F_{o,i}$ is the observed flux, $M_{d}F_{m,i}$ is the model flux of the star,
$M_{d} = \big(\frac{R}{D}\big)^2$ is the scaling factor corresponding to the star (where R is the radius of the star and D is the distance to the star) and $\sigma_{o,i}$ is the error in the observed flux. The number of observed photometric data points N for stars varies from 6 to 25 depending upon their detection in different available filters. The number of free parameters ($N_{f}$) used to fit SED depend on the selected theoretical model. In general, the free parameters are \big[Fe/H\big], log(g) and $T_{eff}$. The radius of the stars were calculated using the scaling factor, $M_{d}$.\\

\begin{figure*}
\centering
\begin{subfigure}{0.5\textwidth}
  \centering
  \includegraphics[height=8.0cm,width=9.0cm]{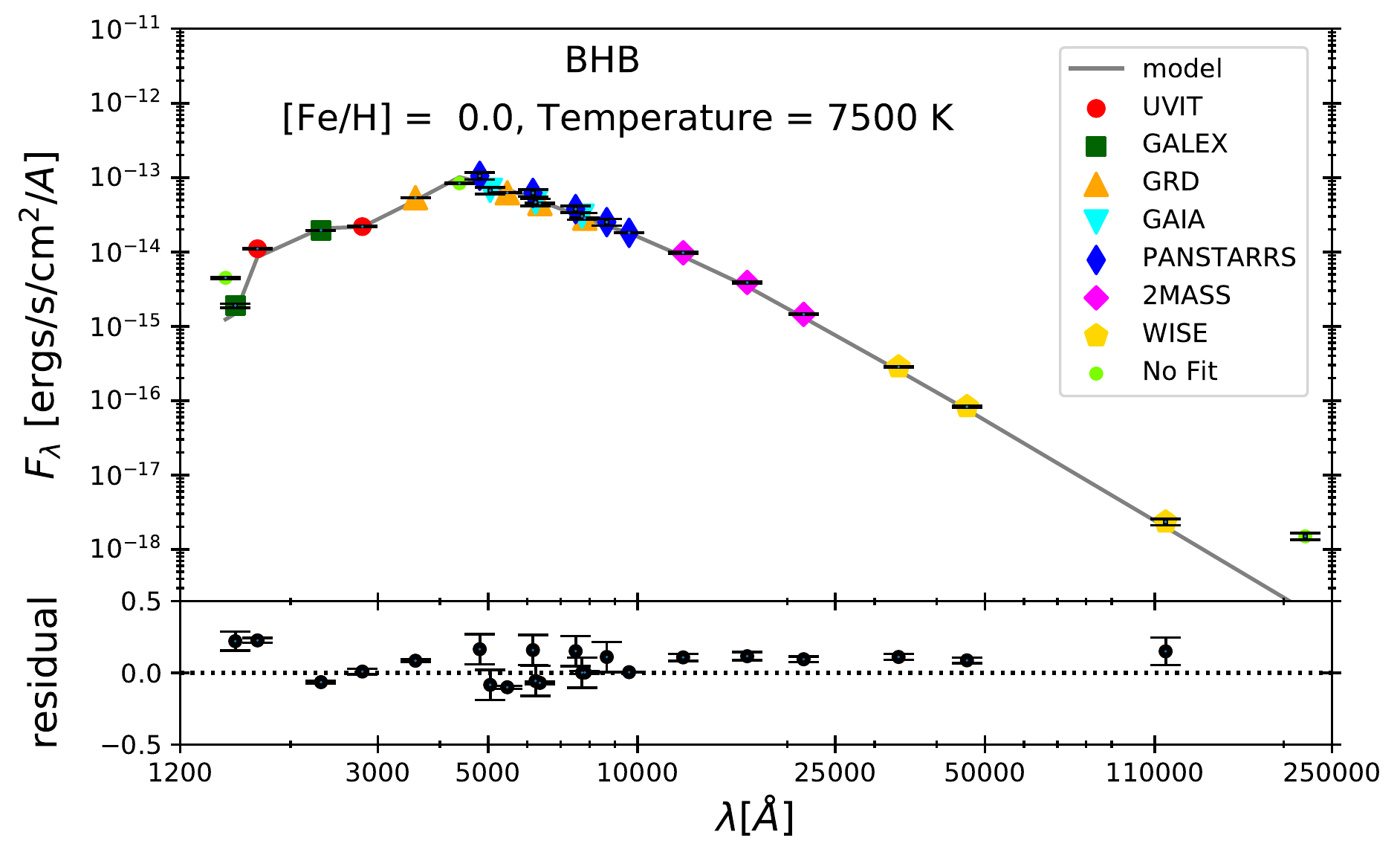}
  \end{subfigure}%
\begin{subfigure}{0.5\textwidth}
  \centering
  \includegraphics[height=8.0cm,width=9.0cm]{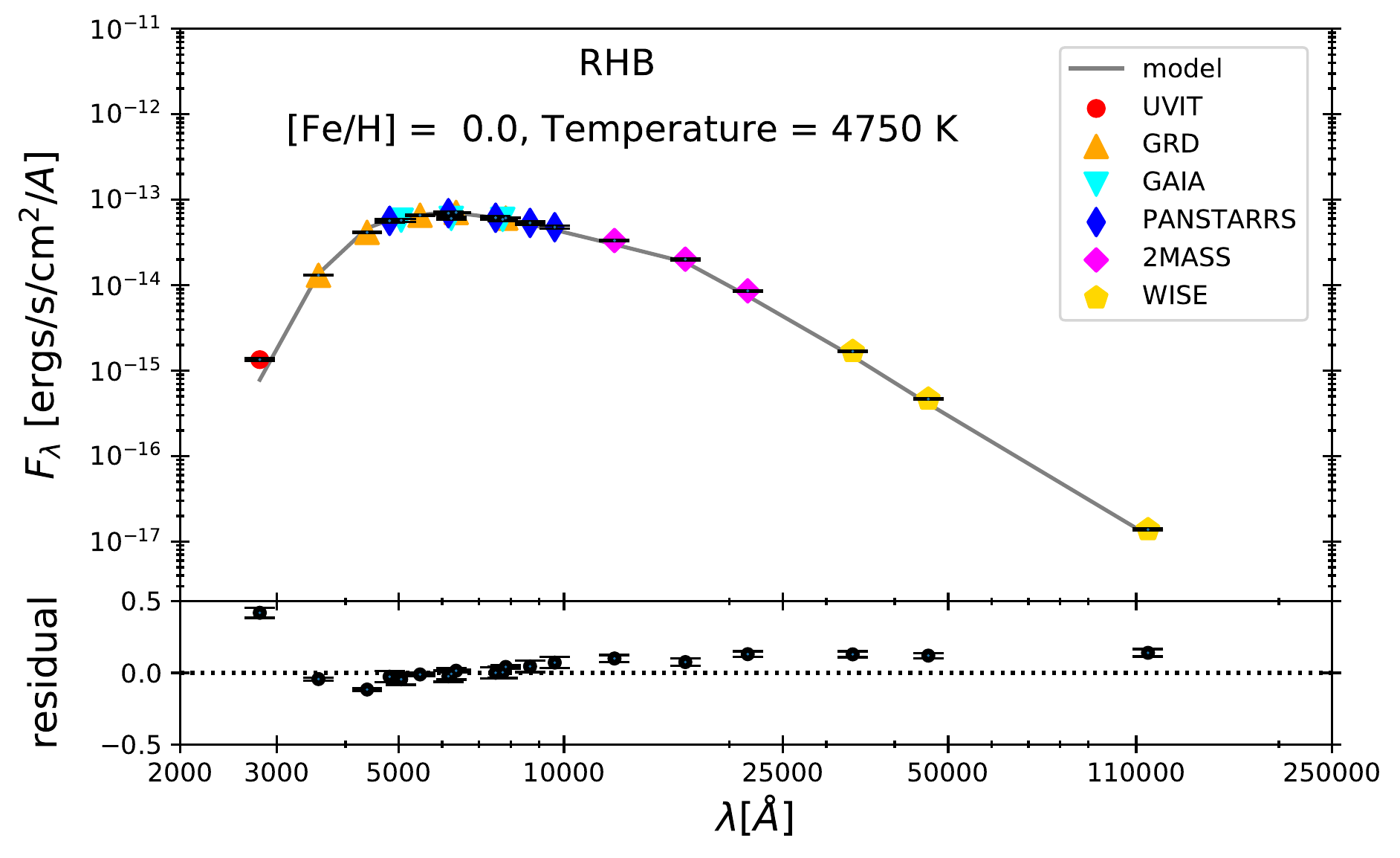}
\end{subfigure}
\begin{subfigure}{0.5\textwidth}
  \centering
  \includegraphics[height=8.0cm,width=9.0cm]{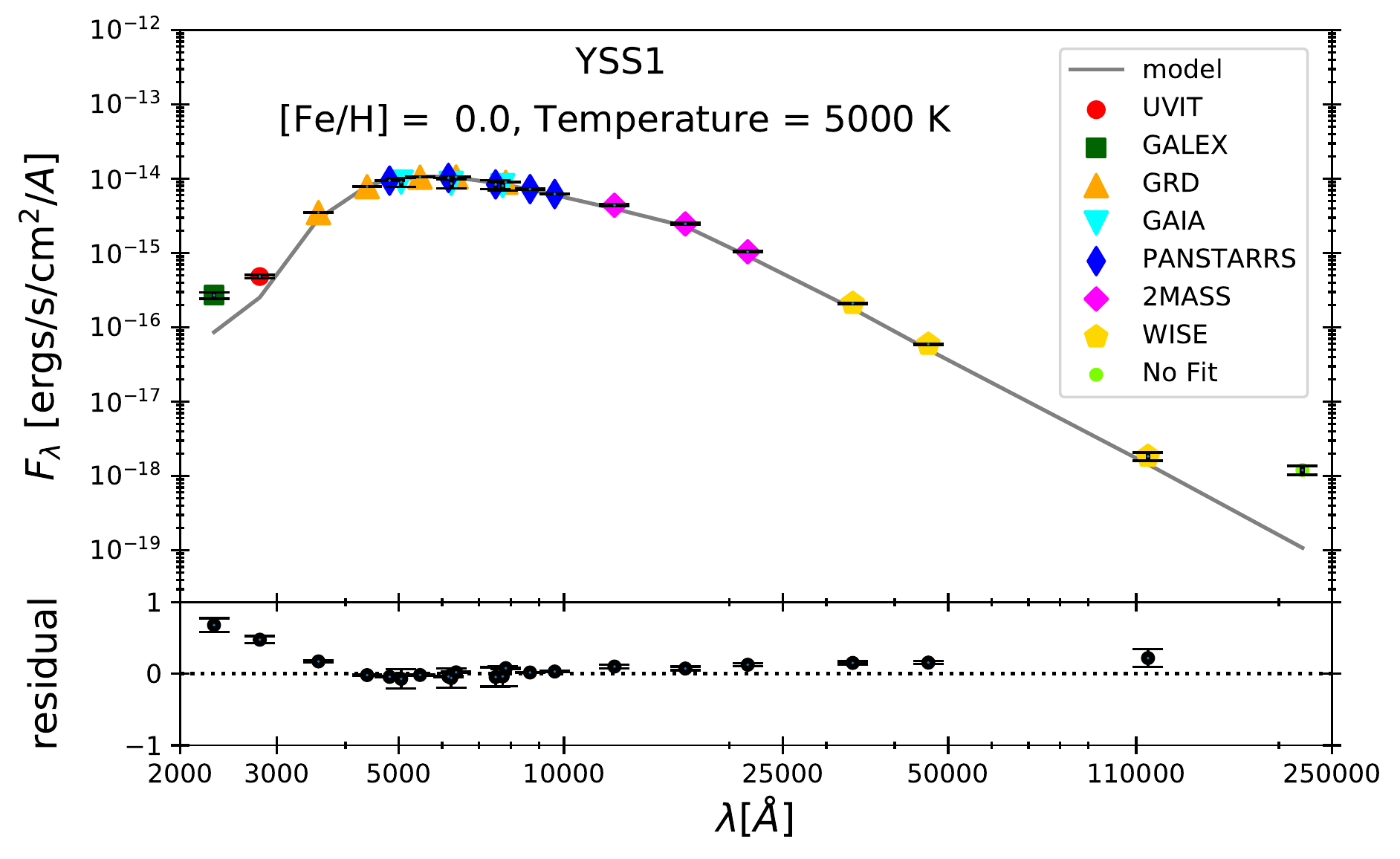}
  \end{subfigure}%
\begin{subfigure}{0.5\textwidth}
  \centering
  \includegraphics[height=8.0cm,width=9.0cm]{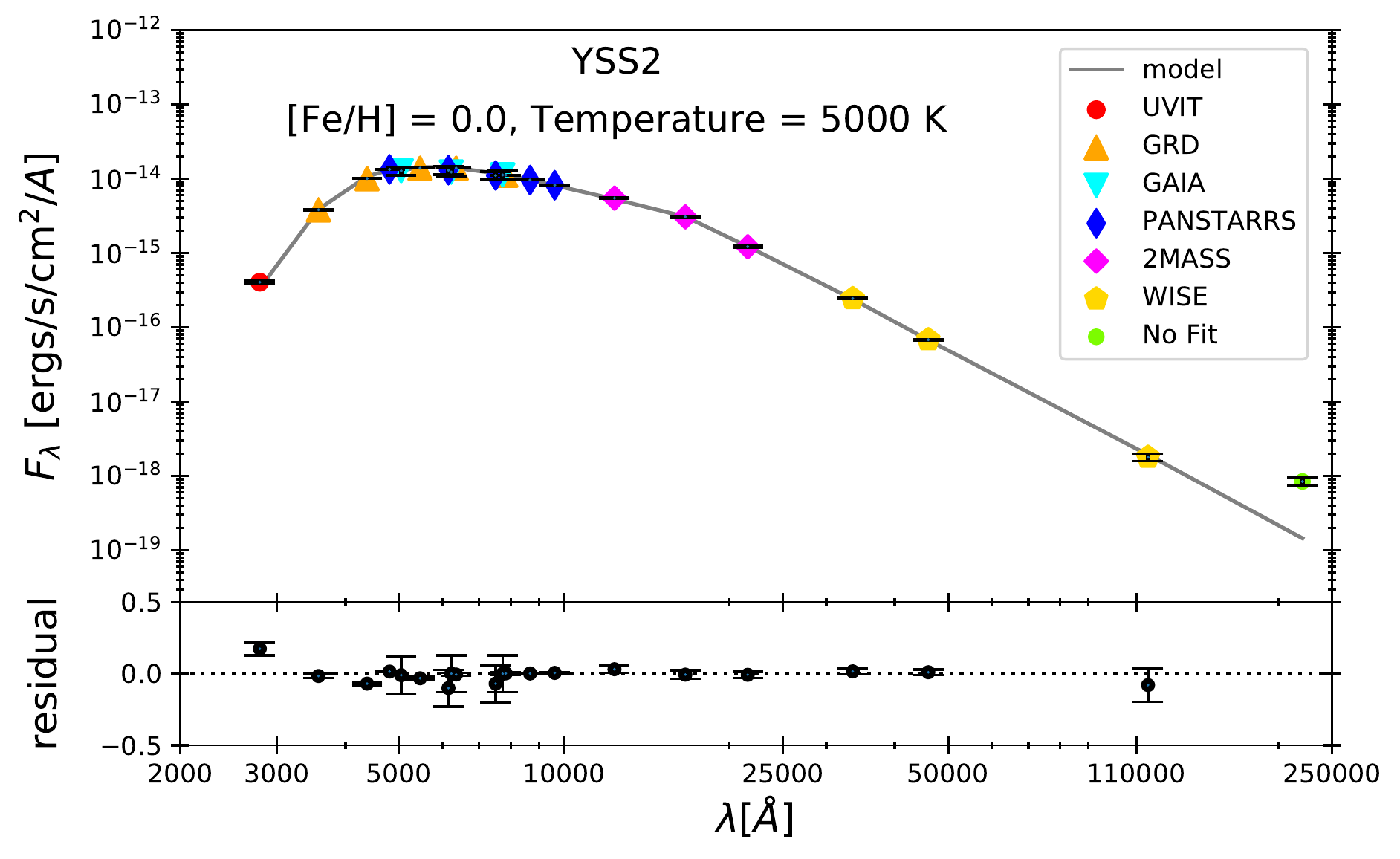}
\end{subfigure}
\caption{Spectral energy distribution (SED) of two HB stars (upper panels) and two YSSs (lower panels) detected with UVIT after correcting for extinction. 
The best fit parameters are displayed in the figure.}
\label{hbseds}
\end{figure*}

\begin{table*}[t]
\small
		\caption{SED fit parameters of bright stars detected with UVIT in this cluster. Column 1 lists the star ID used in this work. Column 2 represents the identification number (WOCS ID) according to Platais {\em et al.} (2003). Columns 3 to 5 show the RA, DEC and model used for SED fit, respectively. Rest of the columns give the estimated values of various parameters along with errors. The last column contains the ratio of the number of photometric data points used for SED fitting and the number of total data points available for fitting.}
    \centering
	\label{sedparam}
	\makebox[0.79\linewidth]
	{
	\begin{tabular}{cccccccccc} 
		\hline
		\hline
		 Star ID & PKM ID & RA (deg) & DEC (deg) & Model &$T_{eff}$ (K) & $\frac{L}{L_{\odot}}$  & $\frac{R}{R_{\odot}}$ & ${\chi}_{red}^2$ & $\frac{N_{fit}}{N_{tot}}$\\
		\hline
		 Subdwarf & WOCS 4918 & 11.96753 & 85.31897 & Kurucz & $21000 \pm 500$ & $6.71 \pm 0.22$ & $0.19 \pm 10^{-3}$  & 14.23 & 19/23\\
		 WD & WOCS 4073 & 10.32108 & 85.23638 & Koester & $20000 \pm 312$ & $0.039 \pm 0.003$ & $0.02 \pm 10^{-4}$ & 0.95 & 6/6\\
		 BHB & WOCS 3856 & 9.77341 & 85.15604 & Kurucz & $7500 \pm 125$ &  $49.13 \pm 7.49$ & $4.13 \pm 0.01$ & 32.92 & 22/25\\
		 RHB & WOCS 5027 & 11.98741 & 85.24889 & Kurucz & $4750 \pm 125$ & $73.83 \pm 3.97$ & $12.56 \pm 0.03$ & 30.85 & 20/20\\
		 YSS1 & WOCS 4705 & 11.34416 & 85.21065 & Kurucz & $5000 \pm 125$ & $10.28 \pm 1.33$ & $4.26\pm 0.01$ & 38.99 & 21/22\\
		 YSS2 & WOCS 4346 & 10.64712 & 85.22385 & Kurucz & $5000 \pm 125$ & $13.63 \pm 1.95$ & $4.94 \pm 0.01$ & 5.6 & 20/21\\
		 \hline
	\end{tabular}
	}
\end{table*}
Kurucz stellar atmospheric models are used to construct SEDs for two HB, one subdwarf, and two YS stars which are bright in NUV band when compared to RGB stars (Castelli {\em et al.} 1997; Castelli \& Kurucz 2003). The model provides the temperatures ranging from 5000-50000 K, logg from 0-5 and metallicity from $-$2.5-0.5 dex. Since this cluster has solar metallicity, we fixed the metallicity \big[Fe/H\big]=0 close to the cluster metallicity. We have given a $T_{eff}$ and logg range from 5000-50000 K and 2-5 dex, respectively for the adopted Kurucz models to fit the SEDs of above mentioned stars. We have combined three UVIT photometric data points with two GALEX, five ground photometry, three GAIA, five PANSTARRS, three 2MASS, and four WISE photometric data points to generate SEDs for UV bright stars. The number of observed photometric data points used for fitting will be equal to or less than the above mentioned data points as some stars are not detected in all the filters. 
 The data points, which were not fitting well to the theoretical model, are also excluded from the fit. 
We have also constructed SED for a possible WD detected with UVIT in both FUV and NUV images using a Koester WD model (Koester 2010; Tremblay \& Bergeron 2010). The free parameters of the Koester model are $T_{eff}$ and logg. The value for the $T_{eff}$ for the Koester model ranges from 5000-80000 K and logg from 6.5-9.5 dex. We have used two UVIT photometric data points along with two GALEX and two Ground photometric data points to fit the SED of the WD. VOSA makes use of Fitzpatrick reddening relation (Fitzpatrick 1999; Indebetouw {\em et al.} 2005) to correct for extinction in observed data points.\\

Best SED fits are obtained for six stars, out of which two are HB, two are YSSs, one is Subdwarf, and one is WD. The SEDs of all the stars are shown in Figures~\ref{seds} and \ref{hbseds}. We can notice in Figures~\ref{seds} and \ref{hbseds} that all the data points are fitted well to the models. 
The estimated values of stellar parameters from the SED fit along with errors are tabulated in Table~\ref{sedparam}. VOSA estimates uncertainties in the effective temperatures as half the grid step, around the best fit value. The high temperature of subdwarf (21000 K) suggests that it belong to the class of subdwarf B (sdB) stars (Heber 1986). The effective temperature of the WD turns out to be 20000 K with a radius 0.02$R_{\odot}$, which confirms the star to a possible WD candidate.\\

In order to check the evolutionary status of HB stars identified with UVIT, we have plotted the theoretical evolutionary tracks using the updated BaSTI-IAC models presented by Hidalgo {\em et al.} (2018). We have selected the model with metallicity close to the cluster metallicity. The evolutionary tracks starting from MS tun-off to the moment when a star has entered to the Post-HB phase are shown in the H-R diagram in Figure~\ref{evotrack}. To show the WD cooling tracks, we used WD models for a 0.5$M_{\odot}$ from Tremblay {\em et al.} (2011). The ZAHB and Post-ZAHB tracks are shown for a mass range from 0.475M$_{\odot}$ to 0.8M$_{\odot}$. The TAHB 
is shown with dash-dotted line in Figure~\ref{evotrack}. For WDs, cooling sequence for a 0.5$M_{\odot}$ DA type WD is shown with dash-dotted black line. The parameters estimated from the SED fit for six stars are plotted in H-R diagram. The stars are marked with same colour as in Figure~\ref{nuvcmds}. We can see in Figure~\ref{evotrack} that HB stars are lying along the HB tracks suggesting that these two stars belong to the HB evolutionary phase. The location of two YSSs on the H-R diagram is near the theoretical RGB sequence. It indicates that these two stars might belong to a giant evolutionary stage. The location of WD on the H-R diagram is near the theoretical WD cooling sequence, implying that the star is likely to be a WD candidate. 

\section{Discussion}
\label{sec:4}
\begin{figure}[t]
\hspace{-0.7cm}
\includegraphics[height=9.0cm,width=9.0cm]{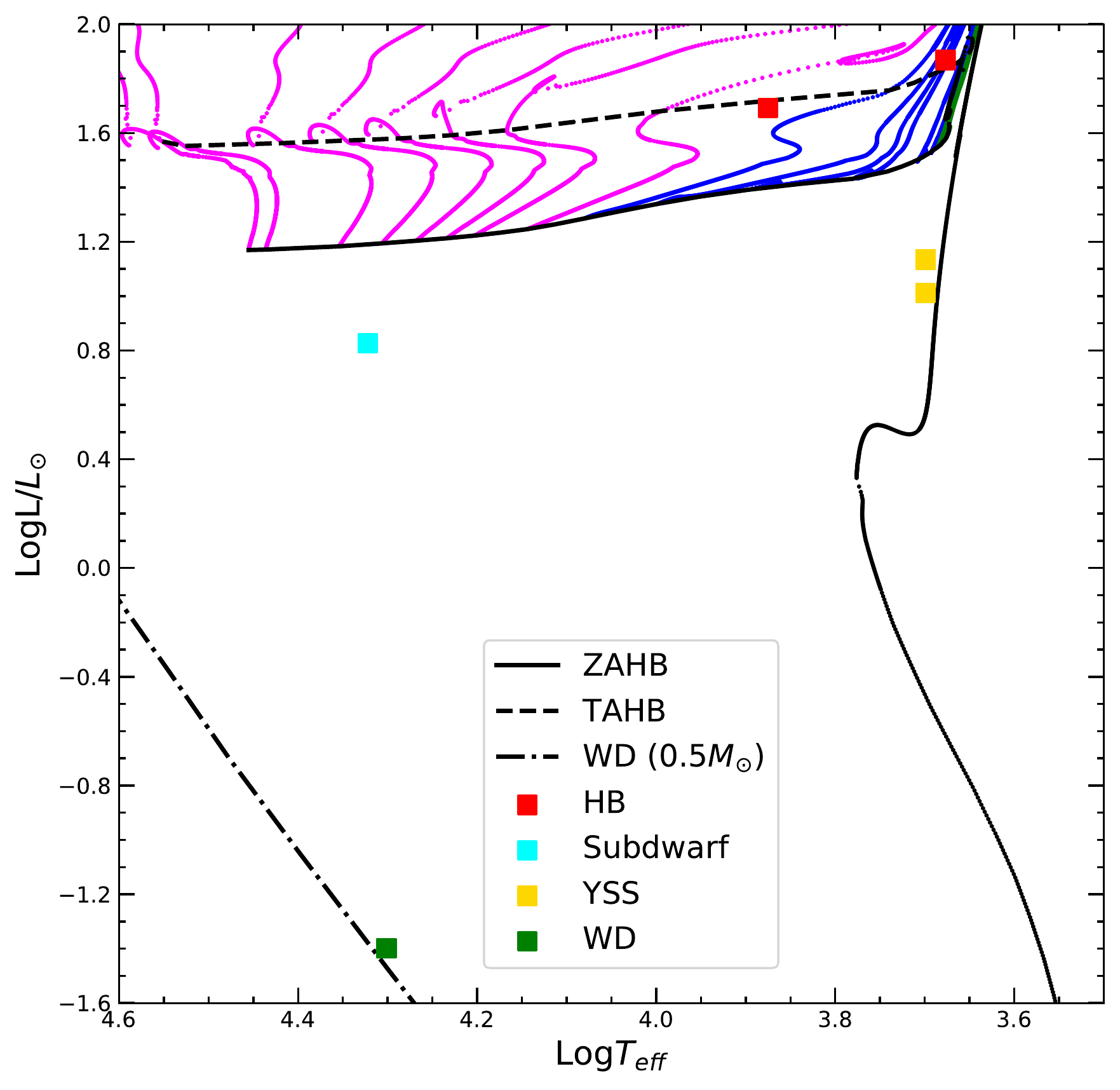}
\caption{H-R diagram of UV-bright stars in NGC 188 compared to theoretical evolutionary tracks. The evolutionary tracks starting from MS turn-off to the moment when a star has entered to the post-HB phase (Hidalgo {\em et al.} 2018) are shown. Along the HB phase, Post-ZAHB tracks span a mass range from 0.475 $-$ 0.8M$_{\odot}$. In the plot, magenta, blue and green colours correspond to the sequences populating the extreme, blue and red parts of the HB. The Black solid and dashed lines indicate the ZAHB and TAHB, respectively. The dash-dotted black line corresponds to cooling track for a 0.5$M_{\odot}$ WD with hydrogen envelope. The SED fit parameters of HB stars, Subdwarf, WD and YSSs identified with UVIT are shown with red, cyan, green and yellow square filled symbols, respectively.}
\label{evotrack}
\end{figure}

In the NUV passband, we have detected 2 HB, 21 BSSs, 2 YSSs, one WD candidate along with MS, MS turn-off, SGB and RGB stars, whereas in FUV, only hot and bright BSSs, WD and HB stars are detected. For comparison with theoretical predictions, we overlaid the CMDs with updated BaSTI-IAC and WD model isochrones generated for respective UVIT and Ground based filters. The UV magnitude distribution of all the detected member stars with UVIT is reproduced well with theoretical isochrones. Out of 21 BSSs detected with UVIT, 15 BSSs are characterised by Gosnell {\em et al.} (2014, 2015) using SED fitting technique. They found hot and young WD companions to 7 BSSs, out of which 5 are detected in our UVIT images. 
7 BSSs binaries are likely formed from mass transfer. Other 6 BSSs are not studied in detail in literature till now. We detected one BSS, which is both FUV and NUV bright. This BSS is a PM member of the cluster, but RV membership is not known. 
Detailed SED analysis is required to check whether the BSS has any hot companion. We plan to characterise the hot companions of the BSSs of this cluster in a separate study and locate them in the H-R diagram (Figure~\ref{evotrack}).\\ 

Out of three UV-bright stars identified with UIT, one star II-91 (numbered by Sandage 1962) is detected in both FUV and NUV images, and here it is designated as a subdwarf. This star was first identified by Sandage (1962), and used for calibration. Later Dinescu {\em et al.} (1996) confirmed its membership based on the proper motion. Green {\em et al.} (1997) performed the spectroscopy of UV-bright stars in open clusters NGC 6791 and NGC 188. They reported that II-91 star is an sdB star and also a spectroscopic binary. Infact, it is a close binary of 2.15 days orbital period (Green {\em et al.} 2004). Landsman {\em et al.} (1998) estimated the effective temperature of II-91 about 30000 K from its UIT magnitude and $m_{152} - V$ colour. Our UV photometry confirms that this star is a UV bright. The T$_{eff}$ measured from SED fitting also confirms it to be an sdB or EHB star.\\

The T$_{eff}$ of one HB star derived from SED analysis is found to be 7500 K, which is in the range of the T$_{eff}$ of a BHB star. For the other HB star, derived temperature is 4750 K, hence likely to be a RHB star in the cluster. The observed UV magnitudes of HB stars match well with theoretical TAHB isochrones, which indicates that these stars are about to evolve off the HB phase, assuming that these are in fact HB stars. Geller {\em et al.} (2008) classified BHB star as a rapid rotator (RR) and binary likely non-member (BLN) as they were unable to measure its radial velocity because of its high rotation. But this star is PM member of the cluster with membership probability more than $80\%$ (platias {\em et al.} 2003). According to Dinescu {\em et al.} (1996), RHB (D719) is one of the brightest giants in NGC 188. Belloni {\em et al.} (1998) identified X-ray sources in OCs M67 and NGC 188 using ROSAT observations. RHB (X29) star is detected as an X-ray source in the cluster. The spectroscopic study of this star (Harris \& McClure 1985) found it to be a fast rotator with rotation velocity $\sim24$ kms$^{-1}$ and also this star exhibits emission in the Ca H and K and H$\alpha$ lines. The absence of radial-velocity variations suggest that this star is a single, rapidly rotating giant, an FK Comae-type star. Our SED fit of RHB star also suggests that it is a single star. From V mag and B$-$V colour, the T$_{eff}$ of this star was estimated to be $\sim$4800 K. Our temperature measurement for RHB star is in close agreement with previous estimations. We suggest that the chromospheric emission of this star might make it bright in NUV region. \\ 

Two YSSs characterised in this work are also RV members, and YSS2 star is previously identified as a giant and single member. Geller {\em et al.} (2008) suggested that YSS1 is a double-lined spectroscopic binary (SB2). Mazur \& Kaluzny (1990) identified YSS1 as a variable star (V11). They also suggested that YSS1 might be an RS CVn-type binary. Gondoin (2005) studied the X-ray sources in NGC 188 using \textit{XMM-Newton} observations, and he identified YSS1 (S18) star as an X-ray source. He also estimated the bolometric luminosity and effective temperature of this star about $\sim$8L$_{\odot}$ and 5110 K, respectively. Landsman {\em et al.} (1997) obtained spectra of the yellow giant S1040 in the OC M67, and found that the star is a single-lined spectroscopic binary. They estimated effective temperature of cool component as 5150 K with a radius 5.1$R_{\odot}$. Our estimation of luminosity ($\sim$10L$_{\odot}$), radius ($\sim$5$R_{\odot}$) and T$_{eff}$  (5000 K) of the YSS1 from SED fitting is close to this value. As YSS1 is a double-lined spectroscopic binary, it might have two components with similar temperature. From SED fitting, we can not separate two components with a similar temperature present in a binary system. The SED fit parameters determined for two YSSs indicate that these two stars may belong to the giant phase. Thus UV emission in the NUV region in the case of YSS1 is likely due to chromospheric activity, probably connected to its binarity and X-ray emission.\\

von Hippel \& Sarajedini (1998) studied the WDs in NGC 188 using the WIYN 3.5-meter telescope at Kitt Peak National Observatory. They identified 9 WD candidates, of which 3$-$6 are expected to be cluster members. Andreuzzi {\em et al.} (2002) identified 28 candidate white dwarfs in the cluster using the data in HST WFPC2 F555W and F814W filters, but their membership was not certain. We have also identified one possible WD candidate in one NUV and one FUV band. We checked with the previous available catalog, but this star is not reported in the earlier studies. 


\section{Summary}
\label{sec:5}
The important results from this study are summarised below:
\begin{itemize}
    \item In this study, we employed UVIT observations on-board \textit{AstroSat} to identify UV-bright stars in the well-known old open cluster NGC 188. We further created the optical and UV-optical CMDs of member stars co-detected using UVIT and Ground-based data in this cluster.
    \item  Stars belonging to the different evolutionary stages such as MS, SGB and RGB  are detected in NUV image, but only hot stars are detected in FUV images.
    \item To compare the observations with theoretical predictions, optical and UV-optical CMDs are overlaid with updated BaSTI-IAC and WD models generated for respective UVIT and ground-based filters. The theoretical isochrones reproduce the features of the CMDs very well.
    \item This study presents the first NUV CMD for this well studied cluster. The CMDs suggest the presence of HB in this cluster, with an RHB, a BHB and an EHB/sdB star populating a temperature range of 4750 K - 21000 K. 
    \item We suggest two YSSs in this cluster, based on their location in the CMDs. YSS1 is found to have excess flux in the UV, may be connected to its binarity and x-ray emission.
    \item We detect a candidate WD from the UV images and is found to have parameters similar to that of a 0.5M$_\odot$ WD.
\end{itemize}

\vspace{1em}

\section*{Acknowledgements}
We are grateful to the anonymous referee for the comments and suggestions which improved the quality of the manuscript. UVIT project is a result of collaboration between IIA, Bengaluru,  IUCAA,  Pune,  TIFR,  Mumbai,  several  centres  of ISRO, and CSA. This publication uses the data from the \textit{ASTROSAT} mission of the Indian Space Research  Organisation  (ISRO),  archived  at  the  Indian  Space  Science  Data Centre (ISSDC). We warmly thank Pierre Bergeron for providing us the WD cooling models for UVIT filters. S. Rani thanks Vikrant Jadhav for useful discussions. This research made use of  VOSA,  developed  under  the  Spanish  Virtual  Observatory  project  supported  by  the  Spanish  MINECO  throughgrant AyA2017-84089. This research also made use of Topcat (Taylor 2005, 2011), Matplotlib (Hunter 2007), NumPy (Van der Walt {\em et al.} 2011), Astropy (Astropy Collaboration {\em et al.} 2018) and Pandas (McKinney 2010).
\vspace{-1em}


\end{document}